# Evaluation of Temporal Formulas Based on "Checking By Spheres"[1]

Wiktor B. Daszczuk
Institute of Computer Science, Warsaw University of Technology
ul. Nowowiejska 15/19, Warsaw 00-665, Poland,
e-mail: wbd@ii.pw.edu.pl
fax (+48 22)6607812

## Abstract

*Classical algorithms of evaluation of temporal CTL formulas are constructed "bottom-up". A formula must be evaluated completely to give the result. In the paper, a new concept of "top-down" evaluation of temporal QsCTL (CTL with state quantifiers) formulas, called "Checking By Spheres" is presented. The new algorithm has two general advantages: the evaluation may be stopped on certain conditions in early steps of the algorithm (not the whole formula and not whole state space should be analyzed), and state quantification may be used in formulas (even if a range of a quantifier is not statically obtainable).*

## I. Introduction

In Institute of Computer Science, Warsaw University of Technology, a software system COSMA is being developed for specification and verification of concurrent systems (hardware, software and co-designed). COSMA is addressed to system level modeling of control dominated systems where the synchronization and cooperation of modules are most important issues.

COSMA is based on CSM automata [1,2,3]. Unlike other formalisms for concurrency, based on interleaving of actions, CSM is based on coincidencies.

For efficient evaluation of temporal formulas in CTL temporal logic [4,5,6,7,8,9,10,11,12,13], an algorithm of finding smallest or largest fixed point of special functional over a state space of a system is often applied [14]. The algorithm of evaluation of temporal CTL formulas is very effective, but it has two disadvantages:

- whole state space must be analyzed for every atomic formula and every operator in a formula;
- whole formula must be analyzed.

In the present paper, a new algorithm of temporal CTL formulas evaluation is presented. The algorithm was especially developed for CSM automata. Both the logic used (QsCTL) and the evaluation algorithm support model checking based on localities, i.e. formulas expressed and evaluated considering behavior of individual automata rather than the global state space. This algorithm may be terminated in very early steps of evaluation on certain conditions. In the new algorithm, only a part of state space may be examined, and some subformulas needed not be evaluated, depending on the progress of evaluation. Of course, in a worst case the whole evaluation must be performed.

In section II, CSM automata are presented along with their state space – Reachability Graph [2]. In section III a QsCTL temporal logic is constructed over Reachability Graph of CSM automata. A notion of characteristic sets, useful in a construction of the algorithm, is introduced in section IV. Section V contains the CBS algorithm for simple temporal formulas. The algorithm is based on Checking By Spheres (CBS) rule. Section VI contains an example of evaluation of simple temporal formula using the CSB algorithm – an error is found, then corrected. The algorithm is generalized to nested formulas in section VII. Time complexity of the algorithm is analyzed in section VIII.

## II. Reachability Graph of a system of CSM automata

A system of CSM automata is presented in [1,2,3]. The general features of CSM automata are:

[1] This work was supported by research project N° 503/G/1032/4030/009 from The Dean of Department of Electronics And Information Technology of Warsaw University of Technology

- component automata are Moore-like (signals are generated in states);
- arcs are labeled with Boolean formulas over input alphabet (if formula is true, the arc may be followed, for example if signal q is active (names of signals are underlined) then the formula $q \vee p \wedge m$ holds);
- arcs leading form a state to the same state are allowed (they are called **ears**);
- automata are complete, i.e. the disjunction of all formulas on arcs leading out of a given state equals true;
- signals (symbols of input alphabet) come from output of automata and from external world;
- special symbols denote the formulas: **1**-always true (an arc may be always followed), **0**-always false (a lack of arc);
- if more than one formula id satisfied on arcs leading out of a given state – the transitions is chosen in non-deterministic way;
- all automata in a system perform always one transition in a lock-step manner (no external clock is required)

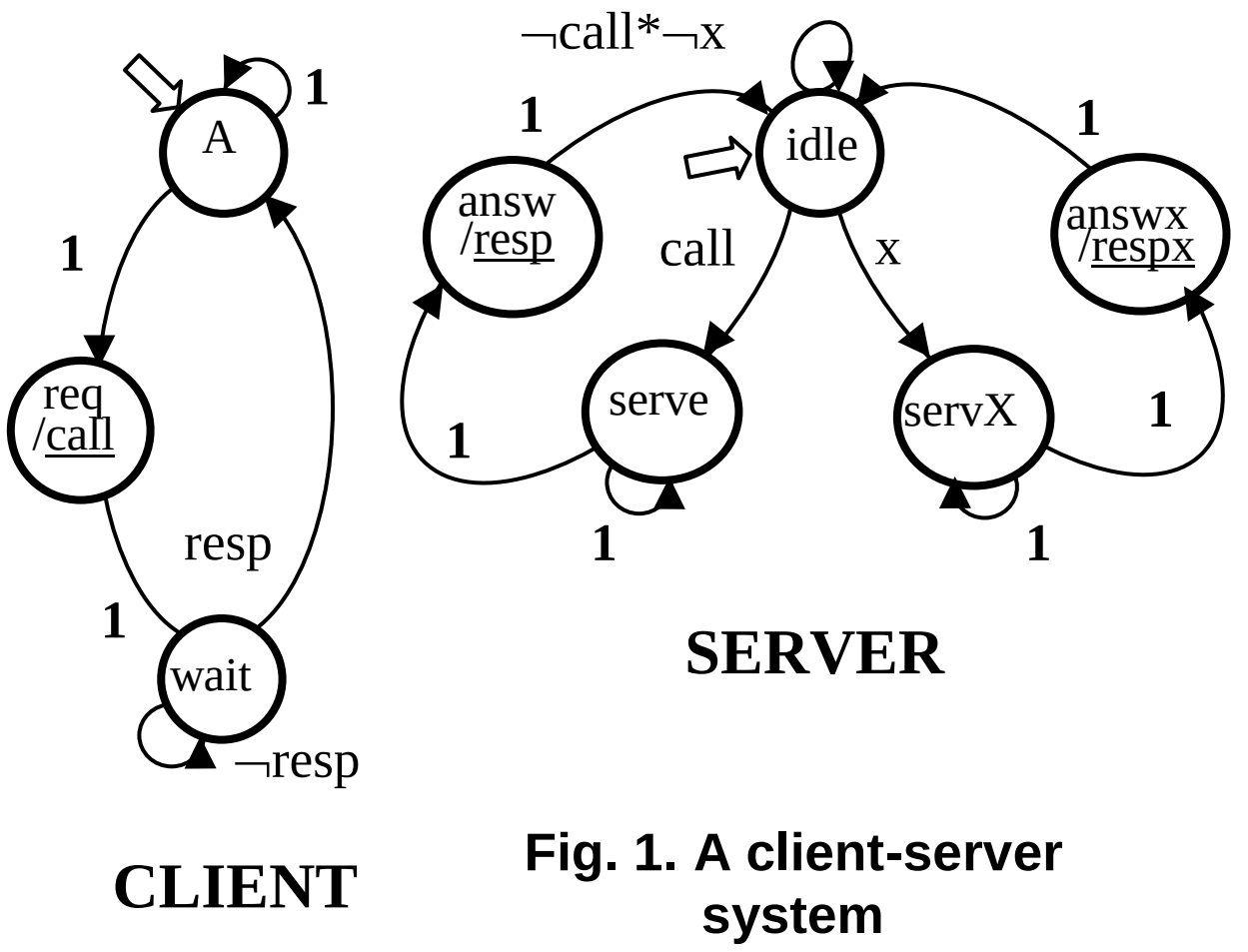

**Fig. 1. A client-server system**

The state space of a system of CSM automata, called Reachability Graph (RG), for a system in Fig. 1 is presented in Fig. 2. States of RG are vectors of states of component automata. Arcs between states are obtained as products of arcs of component automata, i.e. formula on arc of RG is a conjunction of formulas on arcs in component automata.

A set of signals generated in a state of RG is a unions of sets of signals generated in states of components automata. If the state *1* generate signals p,q, and state *3* generates signals q,m, then the state *13* of RG generates signals p,q,m.

The complete algorithm of obtaining RG from component CSM automata is given in [2].

## III. The Temporal Logic QsCTL

The Kripke structure (a base of temporal logic) QsCTL temporal logic constructed over RG is as follows:

- The set of states is simply a set of states of Reachability Graph of CSM automata.
- The succession relation is simply a set of arcs of RG (excluding ears leading out of non-terminal states, therefore a state space is denoted $RG_{-@}$).
- The initial state is an initial state of RG.
- The set of atomic Boolean formulas is:
  - o a signal being generated in states of RG (denoted as a name of the signal in italics),
  - o staying in a given state of RG (denoted ***in*** *s*),
  - o staying in one of a set *S* of states of RG (denoted ***in*** *S*),
  - o staying in a state *s* of RG having a as projection on a component automaton ***a*** given state $s_a$.

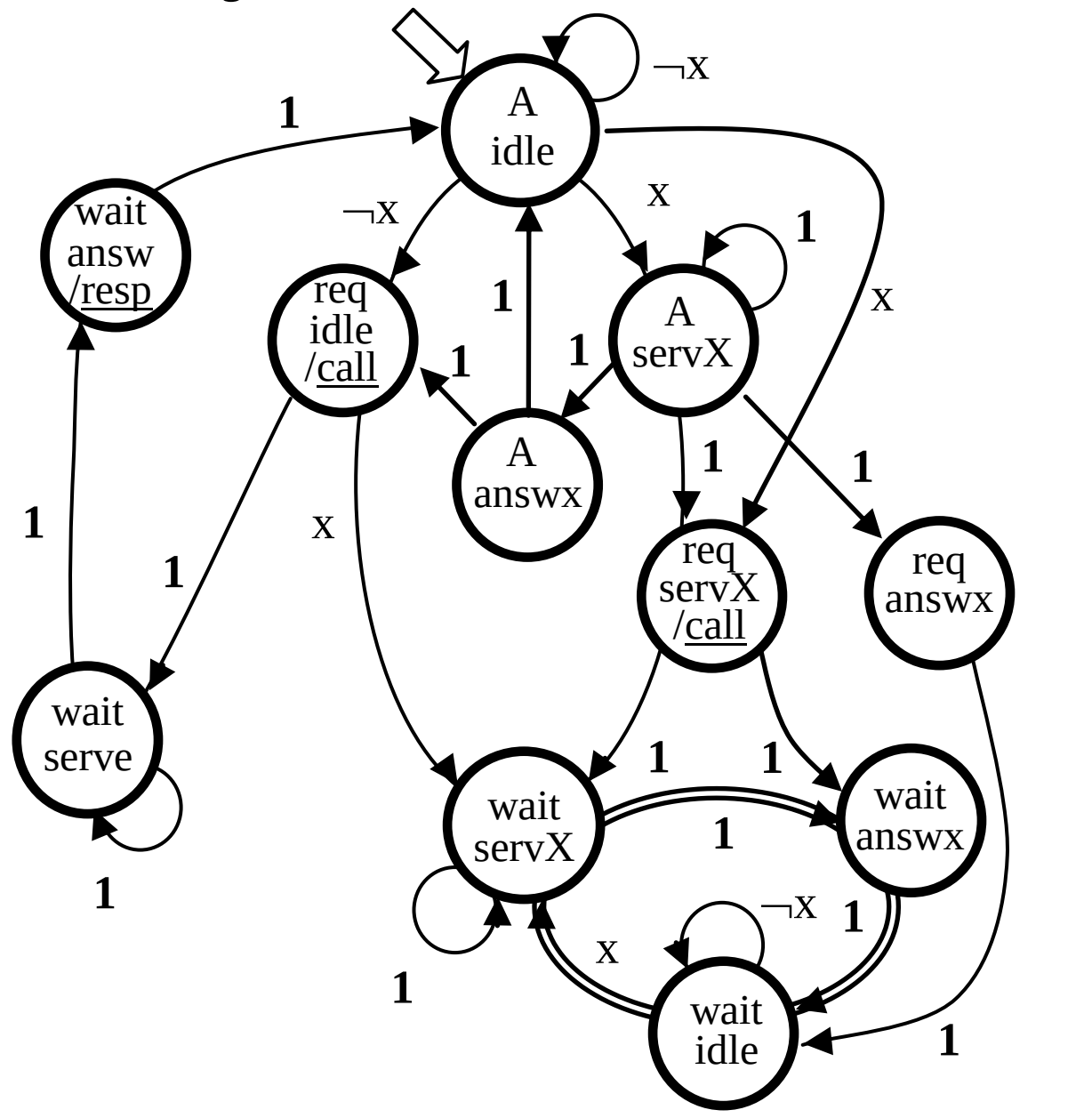

**Fig. 2. RG for the system in Fig. 1**

The modalities used in the temporal logic are as follows:

- **AG**φ - always $\varphi$ in all paths,
- **AF**φ - eventually $\varphi$ in all paths,
- **AX**φ - next $\varphi$ in all paths,
- **A**(φ**U**$_w$ψ) - $\varphi$ until $\psi$ in all paths (weak until),
- $\mathbf{AX_a}$ - next in automaton **a** in all paths (true by definition in a state having a terminal state a as projection on an automaton **a**).

The latter modality is a new one, it simplifies asking questions of correctness in terms of states of component automata rather than in terms of RG.

Additionally, a quantified formulas of a form $\boldsymbol{Q}s\in S; s:\varphi$ ($\boldsymbol{Q}$ is $\forall$ or $\exists$) are allowed. Such formulas are not allowed by classical algorithm of evaluation. In classical algorithm, an evaluation is performed bottom-up (sets of states where atomic formulas are fulfilled are evaluated first, and then embracing formulas). The formula ***in*** *s*, where *s* is a state variable passing through a scope of a quantifier, cannot be evaluated. Checking By Spheres (see section V) is performed top-down and therefore quantified formulas can be evaluated.

For definition of QsCTL temporal logic for systems of CSM automata see [15].

## IV. Characteristic sets

Now, a useful construct of characteristic sets of a given state *s* will be introduced. Characteristic sets are sets of states specific in their reachability relation to state *s*:

The repertoire of characteristic sets for a state *s* is the following:

- a set containing the state *s* itself: *{s}*;
- states reachable from *s*, i.e. a **future** of state *s*; denoted *FUT(s)*;
- states from which the state *s* is reachable, i.e. a **past** of state *s*; denoted *PAS(s)*;
- states reachable from *s* and from which state *s* is reachable (cycles containing state *s*) ; denoted *CYC(s)*; $CYC(s)=FUT(s)\cap PAS(s)$; set *CYC(s)* is a **cyclic future** or **cyclic past** of state *s*;
- states reachable from *s*, from which there is no return to *s*; denoted *END(s)*; *END(s)=FUT(s)-PAS(s)*; set *END(s)* is an **ending future** of state *s*;
- states from which state *s* is reachable, but to which there is no return from *s*; denoted *BEG(s)*; *BEG(s)=PAS(s)-FUT(s)*; set *BEG(s)* is a **beginning past** of state *s*.

The following equations are always fulfilled (*GS* is a set of all states of $RG_{-@}$):

$$BEG(s) = PAS(s) – FUT(s)$$
$$BEG(s) = PAS(s) – CYC(s)$$
$$END(s) = FUT(s) – PAS(s)$$
$$END(s) = FUT(s) – CYC(s)$$
$$CYC(s) = FUT(s) \cap PAS(s)$$
$$GS = PAS(s) \cup \{s\} \cup FUT(s)$$
$$GS = BEG(s) \cup \{s\} \cup CYC(s) \cup END(s)$$
$$GS = BEG(s) \cup CYC(s) \cup END(s) \quad (\text{if } CYC(s)\neq\emptyset)$$
$$GS = BEG(s) \cup \{s\} \cup END(s) \quad (\text{if } CYC(s)=\emptyset)$$
■

## V. The algorithm of Checking By Spheres

The classical algorithm, as said in section I, requires the whole temporal formula to be evaluated (and all its subformulas) in every state of Kripke structure. Moreover, the algorithm presented in [14] checks for all states, reachable and unreachable, and then restricts the result to reachable states only. The algorithm proposed in [15]:

- allows to evaluate only these subformulas that need to be evaluated (for example, if an antecedent of an implication is false, the consequent need not be checked);
- terminates the evaluation as soon as the possible (if the formula $\varphi$ is true in a state *s*, the formula $s:\boldsymbol{AF}\varphi$ is true and does not need to be checked in future of state *s*).

The proposed algorithm is based on **Checking By Spheres** rule. A sphere is a set of states with “distance” from a state *s* of a specific number of arcs. Sphere $SPH_0$ consists of the state *s* itself. Sphere $SPH_1$ contains successors of state *s*. Sphere $SPH_i$ contains states distant from state *s* with *i* arcs, excluding states belonging to spheres $SPH_j$, $j<i$. Fig. 3 illustrates the construction of spheres. Spheres $SPH_0..SPH_3$ contain respectively *1*, *1*, *3* and *1* states.

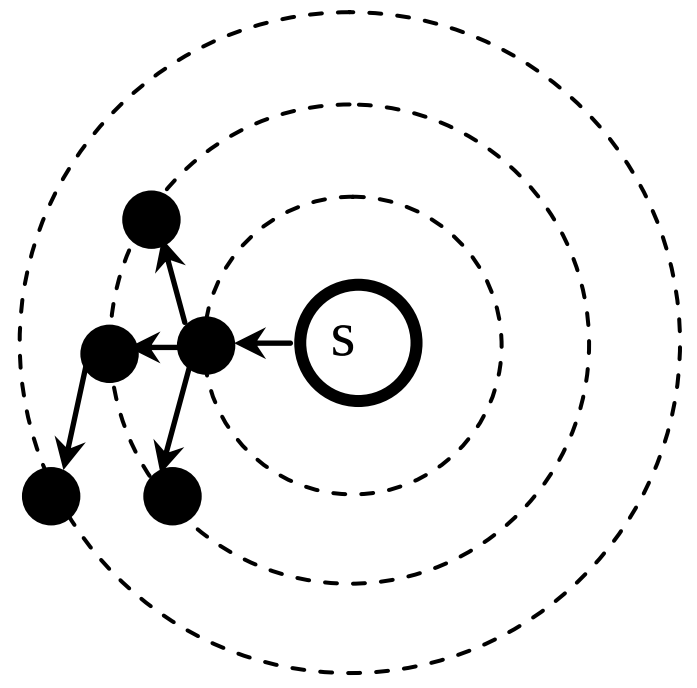

**Fig. 3. Checking By Spheres**

During evaluation of a temporal formula, states in future of a state are checked sphere by sphere. In specific cases of the algorithm, additional conditions are put on states to insert them into spheres. A Checking By Spheres rule (CBS) says that a sphere $SPH_{i+1}$ , $i\geq 0$, is constructed from sphere $SPH_i$ while checking the two conditions:

- if the condition *cond1* holds for a state in $SPH_i$, the algorithm terminates (no other sphere is constructed); the result (*true* or *false*) depends on the state of variable *cond1res*,
- if the condition *cond2* holds for a state $s_j$ in $SPH_i$, the successors of $s_j$ are inserted into sphere $SPH_{i+1}$ (excluding members of spheres *0..i*)

The algorithm terminates with *cond2res* result if there is an empty sphere $SPH_i$, and *cond1* did not appear for any state in $SPH_0..SPH_{i-1}$. A set to search through (to build spheres in it) is denoted *SRC*.

To apply a CBS rule the following elements should be specified:

- *SRC* – a set of states to search through,
- *cond1* – a Boolean condition on which the evaluation terminates prematurely,
- *cond1res* – a Boolean value assigned to the result of evaluation if the evaluation terminates prematurely,

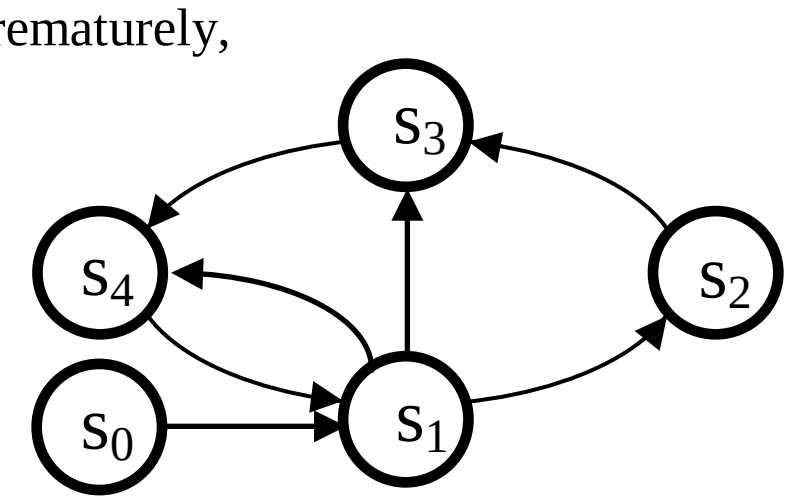

**Fig. 4. Example future of state $s_0$**

- *cond2* – a Boolean condition on which successors of states belonging to $SPH_i$ are inserted into $SPH_{i+1}$,
- *cond2res* – a Boolean value assigned to the result of evaluation if a new constructed sphere is empty and *cond1* did not occur yet.

Nested temporal formula is such that it has a temporal operator (inner operator) in a scope of another operator (outer one). An example of nested temporal formula is:

s : **AX AF in** s

which asks if the future of the state *s* is strictly cyclic (i.e. if after leaving the state *s* there is always a way back to it).

In evaluation of simple temporal formula, the Boolean formula being the argument of temporal operator was evaluated for specific states, according to the rules of algorithm. If a temporal formula is nested, during the evaluation of the outer formula, the inner formula must be evaluated for every state appointed by the algorithm for simple temporal formula. In the above example the formula ***AF in*** *s* should be evaluated for all successors of *s* (states in succession relation with ***r*** with *s*). For other example formula:

s : **AF AG** φ

first, strongly connected subgraphs should be found. For every strongly connected subgraph *CON*, an inner formula must be true either in a single state of *CON*, or during Checking By Spheres in subgraph leading from *s* to *CON*. In every of these states, a formula ***AG*** $\varphi$ is evaluated.

The operation of the algorithm is shown in the case of $\boldsymbol{AU_w}$ operator (weak until). The formula has the form: *s:* $\boldsymbol{A(\varphi\, U_w\, \psi)}$.

In every path one of two conditions must be satisfied:

(a) in every state of a path the formula $\varphi$ is true;

(b) there is a state in a path, in which the formula $\psi$ is true; it must be checked for first such state $s_s$ if for every state between *s* and $s_s$ the formula $\varphi$ is true.

Finding first path in which none of the above conditions is satisfied terminates the evaluation with false result.

As the operator $\boldsymbol{AU_w}$ assumes a „continuity" of holding the truth of the formula $\varphi$ until the formula $\psi$ is satisfied:

- SRC = {s} ∪ FUT(s),
- cond1 = ¬ (φ ∨ ψ), cond1res = false,
- cond2 = φ, cond2res = true.

| Sphere | $s_0$ | $s_1$ | $s_2$ | $s_3$ | $s_4$ |
|---|---|---|---|---|---|
| 0 | • | | | | |
| 1 | | • | | | |
| 2 | | | • | • | • |
| 3 | | ○ | | ○ | ○ |

**Fig. 5. Evaluation of a formula with $AU_w$ operator**

Now the Checking By Spheres will be applied to the evaluation of formula: *s:* $\boldsymbol{A(\varphi\, U_w\, \psi)}$ for RG shown in Fig. 3. Assume that the subformula $\varphi$ holds in every state of RG, but subformula $\psi$ does not hold in any state. The formula is obviously true. Checking By Spheres is illustrated by Fig. 5. Unvisited states have gray background and visited states have white background. States that are candidates to be inserted into next sphere are marked by a dot. If a dot is on gray background, it is inserted into a sphere (black dot). If a dot is on white background, it is not inserted into next sphere because it is visited (white dot). All successors of states in current sphere are candidates to next sphere since in this case $\varphi$ holds continuously, while $\psi$ never. A row (sphere) with no black dot terminates the evaluation.

The implementation of the algorithm in pseudo-code is given in [15].

# VI. Example: A „Client-Server" System

Consider a client-server system presented in Fig. 1. There are three processes: *SERVER*, *CLIENT* and *X* (the latter one not modeled). The *SERVER* performs a service for processes *CLIENT* and *X*. Clients require some services by issuing signals call (*CLIENT*) and x (*X*). The server is in *idle* state until it gets a signal starting a service, then it performs the service (state *serve* or *servX*) and responds to the caller (signal <u>resp</u> for *CLIENT* or <u>respx</u> for *X*)

Fig. 2 contains a Reachability Graph of a system presented in Fig. 1. As an example we will check the formula describing the fact that for every call of a server (signal <u>call</u> in Fig. 1) a reply follows (signal <u>resp</u>).

Let *S* be a set of all states that generate the signal <u>call</u>:

S = {| <u>call</u> |} ■

The requirement to be verified is represented by the following formula:

**AG** ( **in** S ⇒ ( **AF** <u>resp</u> ) )■

It is an abbreviation of formula $s_0$: ***AG* ( *in* S ⇒ ( *AF*** <u>*resp*</u> *) )*, where $s_0$ is an initial state. The outer formula ***AG*** says „always...", therefore the inner (nested) formula ***in*** *S ⇒ (**AF** resp)* must be satisfied in every state. The inner formula is implication, therefore it must be checked if the consequent holds in states where the antecedent is true (in states, in which antecedent is false the consequent is true by definition of implication). The antecedent is true only in states, in which the signal <u>req</u> is generated: states *req_idle*, *req_servX* and *req_answX*. If for any of these states the consequent is false, then the whole formula is false.

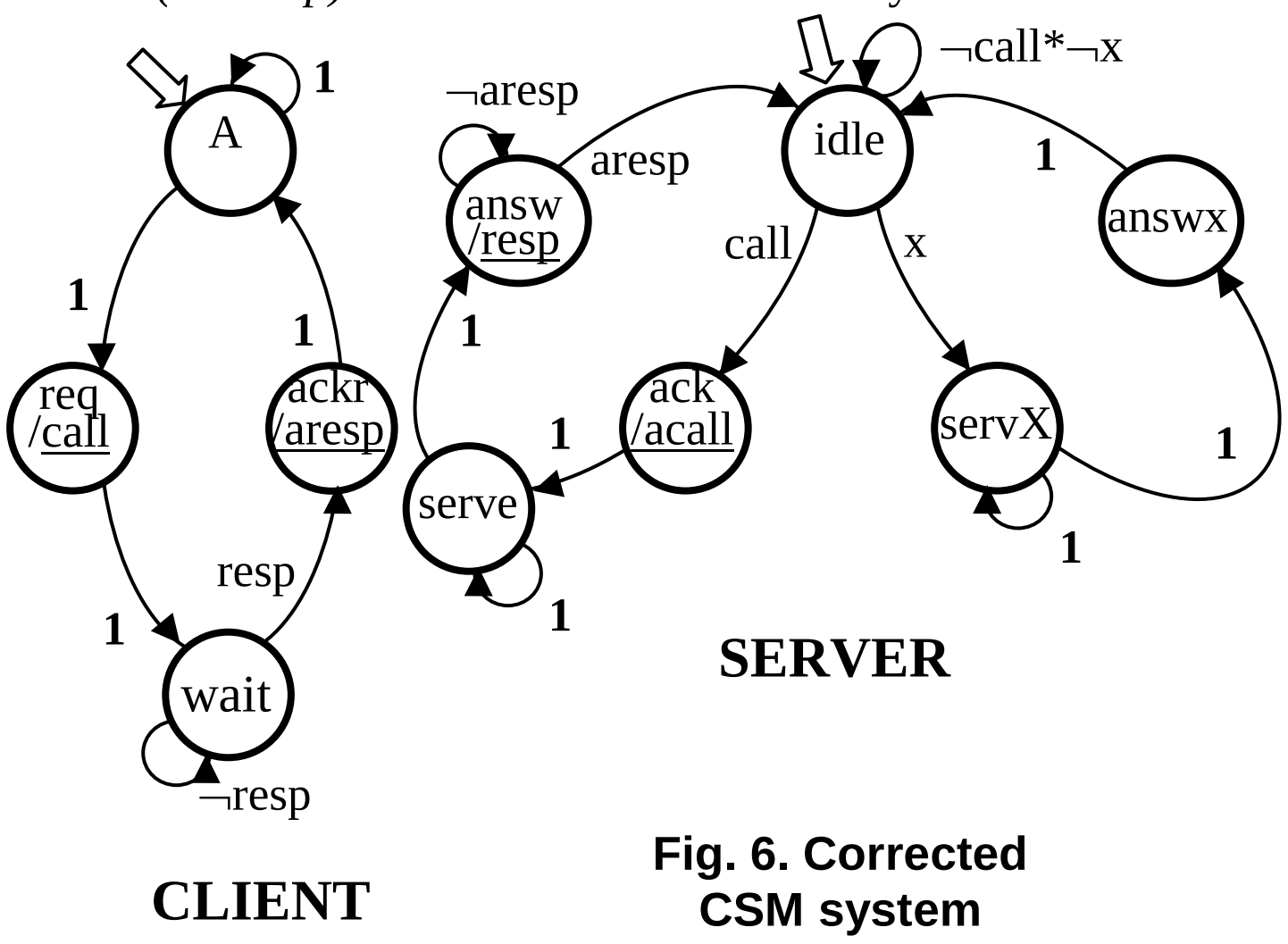

**Fig. 6. Corrected CSM system**

First, let us evaluate the nested formula ***AF*** *resp* for *req_idle*. According to the rules of the algorithm for operator ***AF***, strongly connected subgraphs in a future of *req_idle* should be examined (precise algorithm is given in [15]):

- During Checking By Spheres in the future of *req_idle*, a state $s_i$ out of three states making strongly connected subgraph *CON*($s_i$)= *{wait_servX, wait_answx, wait_idle}* is found (a condition *END*($s_i$)=∅ is fulfilled).
- Then, all states belonging to *CON*($s_i$) are tested for condition *resp* (signal <u>resp</u> being generated in a state). It is done using Checking By Spheres rule.
- None of states belonging to *CON*($s_i$) satisfies the condition, therefore spheres in set *LTC*($s_i$) (states conforming sequences leading to strongly connected subgraph *CON*($s_i$) ) should be checked.
- Sphere $SPH_0$ consists of *req_idle* only. The condition of negative evaluation (*¬resp* and a successor of *req_idle* belongs to *CON*) is true, so the conclusion is that the formula *req_idle* : ***AF*** *resp* is false.
- Therefore, the outer formula ***AG* ( *in*** *S* ⇒ ( ***AF*** <u>*resp*</u> *) )* is false.

The evaluation of the formula shows that the system shown in Fig. 1 does not work properly (the server does not respond to some calls sent by the client) as a result of the lack of acknowledgments. Introduction of acknowledgments requires that the client must keep the call (signal <u>call</u>), until the server acknowledges it. We introduce the signal <u>acall</u> for this purpose. Similarly, the signal <u>acresp</u> will acknowledge the signal <u>resp</u>. The modified system from Fig. 1 is presented in Fig. 6. The new RG for a system in Fig. 6 is presented in Fig. 7.

Again, a nested formula ***in*** *S ⇒ (**AF** resp)* must be satisfied in every state, thus the subformula ***AF*** *resp* should be evaluated for *req_idle*, *req_servX* and *req_answX*. This time, a future of *req_idle* is purely cyclic (*END(req_idle)=∅*), therefore the set *CON=CYC(s)= FUT(s)* will be Checked By Spheres ($\varphi$ = *resp*):

- SRC = FUT(s)
- cond1 = resp, cond1res = true,
- cond2 = true, cond2res = false.

Indeed, there is a state *conf_answ* satisfying *resp*, therefore the formula is true.

The conclusion for the other two states (*req_ack* and *req_answx*) is the same, because a future of every of these states is purely cyclic and the state *conf_answ* belongs to the future of every of the states. The outer formula is positively evaluated: for every call the server sends a response.

## VII. Complexity Of The Algorithm

**Simple formulas.** Let $RG_{-@}$ has N states and M arcs (in worst case, when a graph is a clique, $M=N^2$). The rule of Checking By Spheres is used in every case of the algorithm. During the analysis of a sphere, successors of states form the next sphere. The arcs of $RG_{-@}$ are followed while constructing and checking spheres. Therefore a complexity is O(N*M), result is identical to traditional algorithm [14].

If many operators take place in a formula (number of operators is B), but they do not nest, the complexity multiplies by number of operators, as in classical algorithm O(N*M*B).

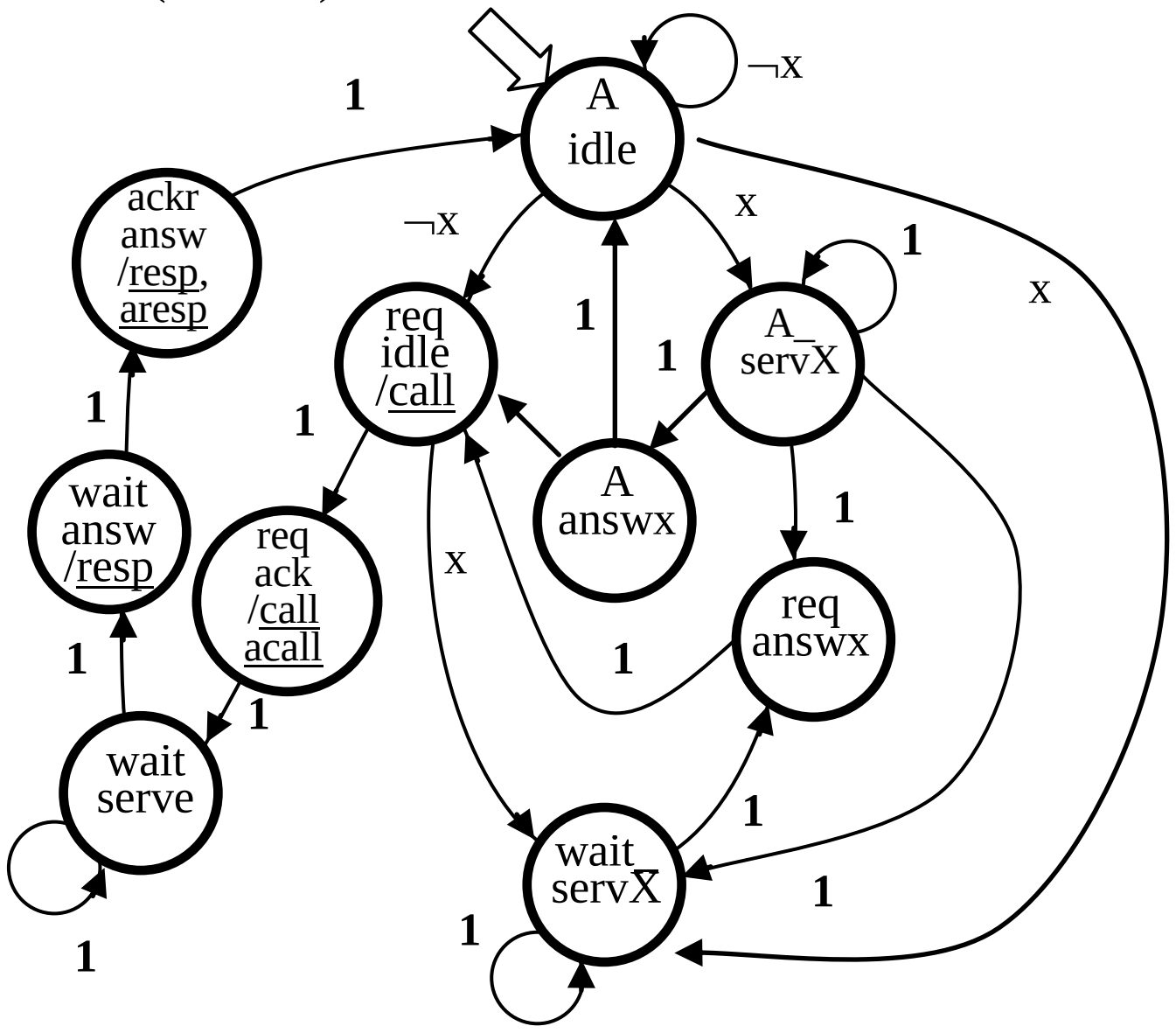

**Fig. 7. RG for CSM system in Fig. 6**

**Nested formulas.** As nested temporal formulas are tested recursively, it may sometimes require a formula to be evaluated for a given state many times (for example ***AG AF*** $\varphi$ - a given state *s* may be in future of many states, therefore the formula $\varphi$ will be evaluated many times for state *s*). In general, it leads to the complexity $O((N*M)^K)$, where K is number of directly nested pairs of operators. It is much worse than for traditional algorithm [14], where the state space is examined exactly as many times as the number of operators take place in a formula, regardless if they nest or not (O(N*M*K)).

This disadvantage can be avoided by constructing a parsing tree of a formula, and storing in every node of the tree information for which states a formula has been checked and for which it is true. The size of sets may be as large as GS (the number of states in RG), but ROBDD representation [14,16,17] can provide a compact representation of sets of states. The idea is discussed in detail in the description of the algorithm for checking using ROBDD [18].

**State quantifiers.** Generally, a set being a scope of a quantifier may not be dynamically obtainable (for example, *FUT(s)*). Therefore, a complexity must be multiplied by N*M (checking state by state in the set, calculating the set in the same time). If a state variable *s* is used in the inner formula, it has some consequences on the manner of evaluation. The values of subformulas in states, stored in parsing tree, must be "forgotten" each time *s* changes its value. As for nested operators, this leads to complexity $O((N*M)^{L+1})$, where L is the number of state variables passing through scopes of nested quantifiers, that have state variables used in quantified formulas.

**Potential advantages.** For the cost of the above complexity, the designer is offered new possibilities of temporal model checking:

- Usage of quantified state formulas simplifies asking many questions.
- Only a part of full state space is often examined during evaluation of formulas (the evaluation terminates as soon as the result is unambiguously calculated – the condition *cond1* is true for any checked state in *SRC* set).
- Some subformulas need not be evaluated at all or for specific states (for example while checking the formula $\boldsymbol{A(\varphi U_w \psi)}$ when $\psi$ is true, or while checking the formula $\psi \Rightarrow \varphi$ when $\psi$ is false, the value of formula $\varphi$ is not important).

Although savings resulting from early termination may be substantial, they cannot be predicted because they depend strongly on the shape of RG and values of atomic Boolean formulas in states. The form of the checked formula is also important.

## VIII. Conclusions

In the paper, a new concept of top-down evaluation of temporal formulas, named "Checking By Spheres" is presented. The idea is addressed to temporal checking of systems modeled by means of CSM automata. Such approach allows to obtain new features:

- Evaluation may be terminated prematurely, if certain condition occurs (not the whole formula and not whole the state space is then analyzed). Therefore, some parts of state space and some subformulas may be omitted during evaluation.
- State quantification may be used in formulas, even if a range of a quantifier is not statically obtainable. A variable passing through a scope of a quantifier may be used inside a formula.

The disadvantage of the new algorithm is its complexity in worst case. It should also be noted that the evaluation of temporal formulas is one of the last stages of verification. The temporal model checking is preceded by calculating of RG and converting it to $RG_{-@}$. The problem of obtaining a state space is NP, but it is a problem of **any** finite state verification method. Generally, the complexity of calculating may be overcome by:

- usage of efficient and concise representation of data structures (ROBDD) and effective procedures and tools for ROBDD,
- hierarchical methods of representation and checking of concurrent systems,
- compositional model checking,
- state space reduction techniques,
- combined methodology: finite state approach plus theorem proving.

## Acknowledgment

I would like to thank Jurek Mieścicki for his help in preparation of this paper.